\documentstyle[11pt,aaspp4,epsf]{article}

\def\kpc{{\rm\,kpc}}

\def\spose#1{\hbox to 0pt{#1\hss}}
\def\lta{\mathrel{\spose{\lower 3pt\hbox{$\mathchar"218$}}
     \raise 2.0pt\hbox{$\mathchar"13C$}}}
\def\gta{\mathrel{\spose{\lower 3pt\hbox{$\mathchar"218$}}
     \raise 2.0pt\hbox{$\mathchar"13E$}}}
 
\def\Re{\mathop{\it Re}\nolimits}
\def\Im{\mathop{\it Im}\nolimits}
 
\def\pdrv#1#2{{\partial #1 \over \partial #2}}
\def\drv#1#2{{d #1 \over d #2}}
\def\tdrv2#1#2{{d^2 #1\over d{#2}^2}}

\newcommand{\half}{{\textstyle{1\over2}}}

\def\bfV {{\bf V}}
\def\bfr {{\bf r}}
\def\bfv {{\bf v}}
\def\bfk {{\bf k}}
\def\bfOmega {{\bf\Omega}}
\font\gkvec=cmmib10
\def\bfomega {\hbox{{\gkvec\char33}}}
\def\bfw {{\bf w}}
\def\bfI {{\bf I}}

\begin{document}

\title{Transmission and amplification of disturbances by massive
	halos}

\author{Chigurupati Murali}

\affil{Canadian Institute for Theoretical Astrophysics, McLennan Labs, 
University of Toronto,\\ 60~St.\ George St., Toronto M5S 3H8, Canada}

\begin{abstract}
We study how massive halos respond to perturbations.  Through
numerical solution of the coupled linearized Boltzmann-Poisson
equations, we find that halos can transmit and amplify disturbances
over large distances within galaxies.  In particular, as Weinberg has
noted, the halo provides an excellent medium for transmitting
disturbances from the outer regions of galaxies to the inner regions.
The dipolar response typically dominates and, in some cases, is very
weakly damped.  The monopole and quadrupole response can also make
significant contributions.  Overall, the results support the notion
that disk structure can be excited by transmission of noise from the
distant halo.  The strong dipolar response suggests that halos play a
role in producing lopsided disks.  These results also suggest that,
during formation from initial collapse, halos do not settle rapidly
into static equilbrium; instead, they may continue to ring and
interact with other embryonic components of proto-galaxies.
\end{abstract}


\section{Introduction}
Galaxies typically inhabit noisy neighborhoods.  Noise-- or
irregularites in the gravitational field-- usually appears in the form
of large satellites in the outskirts of galaxies (e.g. Zaritsky \&
White 1994).  It is also expected in the form of continuing infall and
inhomogeneity in the dark matter distribution.  Yet most of this noise
appears to lie at rather large distances.  We might suppose,
therefore, that it does not disturb the tranquility of the inner
regions of galaxies, particularly galactic disks.

However, massive halos dominate most galaxies.  They extend to large
distances and, in some sense, provide a medium which potentially
fosters communication between spatially separated regions, as
Lynden-Bell (1985) first pointed out.  Recent work by Weinberg
(1995,1998) elucidates the dynamical behavior of this connective
medium: the halo can respond to forcing by a satellite such as the LMC
and efficiently transmit disturbances into the inner regions of
galaxies.  Consequently the apparent spatial isolation of inner
regions is balanced by the halo's ability to transmit disturbances:
the centers of galaxies seem to hear the noise, loud and clear.

Given the variety and ubiquity of disk structure, this might not come
as a surprise.  Galactic disks prominently display warps,
lopsidedness, bars, spiral arms and other asymmetries.  Theoretical
work on disk structure shows that individual features are short-lived
and must be continuously re-excited.  Neverthess, it has been
difficult to pinpoint the culprit since many candidate disturbers such
as the LMC lie too far from the disk to excite structure through
direct tidal forcing.  Halo transmission and amplification seems to
provide at least one necessary mechanism for continuously exciting
disk structure (see Nelson \& Tremaine 1995 for a comprehensive review
of disk dynamics).

Unfortunately, it is difficult to study the response of halos to
moderate amplitude perturbations.  Ideally we could simply perform
simulations.  However, current N-body techniques suffer from
resolution limitations due to particle number, drift in orbit
integration and, in the case of expansion codes, inability to conserve
momentum.  These problems suppress linear collective effects which we
expect to dominate the response in this regime.

An alternative approach, known as the `matrix method', was originally
developed to calculate dispersion relations of disk systems by Kalnajs
(1977) and has been adapted by Weinberg (1989) to study spherical
systems.  The original implementation is best suited for stability
analysis and determining the growth rates of unstable modes
(e.g. Weinberg 1991; Saha 1992) because the dispersion relation is
defined in the upper-half of the complex frequency plane.  Recently,
Weinberg (1994) modified the method to identify modes in stable
systems using elegant numerical techniques to continue the dispersion
relation into the lower half-plane.

In the present work, we further modify the matrix method in order to
calculate the evolution of a perturbed system in time.  In essence, we
solve the coupled linearized Boltzmann-Poisson equations as an initial
value problem to determine the linear response of the model system.
Our approach is well suited to the study of stable galaxy response
because the time-dependence and relative amplitudes of excited modes
are uniquely determined for arbitrary external perturbations.
Weinberg (1989,1995,1998) has taken a similar approach in calculating
the response in the asymptotic limit of a periodic perturbation.

Herein we present the results of calculations using this method.  Our
results support the findings of Weinberg (1994,1995,1998) and
demonstrate that halos can amplify perturbations and absorb energy
into weakly damped oscillatory modes.  Perhaps as importantly, the
results demonstrate the need for high precision numerical calculations
to study the subtle effects which determine the structure and
evolution of galaxies.  The outline is as folllows.  We first derive
the modification of the matrix method in \S\ref{sec:formalism}.  We
then discuss the numerical implementation of the method and the
battery of tests used to verify its accuracy in
\S\ref{sec:implementation}.  The first section on results,
\S\ref{sec:point_perturber}, details the response of King models to
point-like perturbations.  Although not presented in a direct
astrophysical context, these results demonstrate the variety of
effects arising in the models and provide a convenient framework for
categorizing the physical behavior.  In the second section on results,
\S\ref{sec:realistic}, we present an example of a fly-by to illustrate
how effects arise in an astrophysical context.  Several possibilities
for further investigation are discussed in \S\ref{sec:discussion}.

\section{Formalism}
\label{sec:formalism}
\subsection{Hamiltonian}
We choose coordinates centered always on the unperturbed (monopole)
potential.  This introduces an {\it indirect term} into the
Hamiltonian due to the acceleration of the reference frame.  The
resultant Hamiltonian then only describes the internal reaction of the
unperturbed system to the perturbation.  For tidal encounters, the
indirect potential directly cancels the $\ell=1$ component of the
perturber's gravitational potential.  However, for interpenetrating
encounters, the term appears explicitly.

The full Hamiltonian is
\begin{equation}
H=\half \vert \bfv \vert^2+\Phi_0(\vert\bfr\vert)+
\Phi^s(\bfr,t)+\Phi^e(\bfr,t) + 
\drv{\bfV_c}{t}\cdot\bfr\equiv H_0 +\Delta H.
\end{equation}
The indirect term is given by the linear potential
$d\bfV_c/dt\cdot\bfr$, where $d\bfV_c/dt$ is the mean acceleration of
the unperturbed system.  Defining the total potential perturbation
$\Phi_1=\Phi^s+\Phi^e$ as the respective sum of the response and
direct external potentials, we can write
\begin{equation}
\drv{\bfV_c}{t}=-{1\over M_c}\int d\bfr\rho_0\pdrv{\Phi_1}{\bfr}
	={1\over M_c}\int d\bfr\rho_1\pdrv{\Phi_0}{\bfr},
\end{equation}
where $\rho_1=\rho^s+\rho^e$ is the respective sum of response and
external source densities.  The indirect term introduces a small
modification of the matrix equation for the $\ell=1$ component of the
response.  In these coordinates, the response contains no barycentric
shift.

\subsection{Linear theory and matrix method}
\label{sec:linear}

	The matrix method for determining the stability of stellar
systems was originally developed by Kalnajs (1977).  Our derivation is
related to Weinberg's (1989) adaptation of the matrix method to
spherical systems.

	We consider the effect of weak perturbations on an equilibrium
stellar system with distribution function $F({\bf I})$ (hereafter DF),
where the vector ${\bf I}$ denotes the actions associated with
unperturbed orbits in the potential.  Jeans' theorem states that the
equilibrium DF does not depend on the angle variables ${\bf w}$
(Binney \& Tremaine 1987).  The evolution of the perturbed DF $f({\bf
I},{\bf w})$ is described by the linearized Boltzmann equation
\begin{equation}
\pdrv{f}{t}+\pdrv{H_0}{\bfI}\cdot\pdrv{f}{\bfw}-
	\pdrv{F}{\bfI}\cdot\pdrv{\Delta H}{\bfw}=0,
\label{eq:LBE}
\end{equation}
where $\Delta H$ is the perturbing Hamiltonian defined above.  The
response potential $\Phi^s$ is related to $f$ through Poisson's
equation
\begin{equation}
\nabla^2\Phi^s=4\pi G\int f d\bfv\equiv 4\pi G \rho^s.
\label{eq:poisson}
\end{equation}	
In the non-self-gravitating approximation, one sets $\Phi_1=\Phi^e$.

	The perturbed quantities, $f$ and $\Delta H$, may be expanded
in Fourier series in the action-angle variables of the unperturbed
system (Goldstein 1980; Tremaine \& Weinberg 1984).  Following their
convention, we define $I_1$ as the radial action, $I_2$ as the total
angular momentum and $I_3$ as the z-component of the angular momentum.
Motion in the unperturbed Hamiltonian $H_0({\bf I})=\half
\vert\bfv\vert^2+\Phi_0(\vert\bfr\vert)$ is given by $\bfI=$ constant,
$\bfw=\bfomega t+\bfw_0$.  In particular, the Fourier expansion of the
perturbation becomes
\begin{equation}
\Delta H({\bf r},t)=\sum_{\bfk} \Delta H_{\bfk}(\bfI,t) \exp(i\bfk\cdot\bfw),
\end{equation}
where
\begin{equation}
\Delta H_{\bfk}(\bfI,t)={1\over (2\pi)^3}\int d\bfw \Delta H(\bfr,t)
	\exp(-i\bfk\cdot\bfw).
\label{eq:Phi1k}
\end{equation}
Tremaine \& Weinberg (1984) give expressions for the action-angle
series of a multipole potential expansion: we introduce these formulae
below.

	Substituting into equation (\ref{eq:LBE}) yields the relation
\begin{equation}
\pdrv{f_{\bfk}(\bfI,t)}{t}+i\bfk\cdot\bfomega f_{\bfk}
	=i\bfk\cdot\bfomega\drv{F}{E}\Delta H_{\bfk}(\bfI,t),
\label{eq:LBEk}
\end{equation}
since $\partial H_0/\partial \bfI={\bfomega}$ defines the vector of
stellar orbital frequencies and $\partial F/\partial\bfI=\bfomega
dF/dE$ for an isotropic DF.  We can construct the inhomogeneous
solution to equation (\ref{eq:LBEk}) from the homogeneous solution
(right-hand side zero); therefore
\begin{equation}
f_{\bfk}(\bfI,t)=i\bfk\cdot\bfomega\drv{F}{E}\int_{-\infty}^t dt' 
	\exp[i\bfk\cdot\bfomega (t'-t)]\Delta H_{\bfk}(\bfI,t').
\label{eq:nsgsol}
\end{equation}
Summing over $\bfk$ and integrating over velocities yields a formal
expression for the perturbed density:
\begin{equation}
\rho^s=\int d\bfv\sum_{\bfk}\exp(i\bfk\cdot\bfw)(i\bfk\cdot\bfomega)
	\drv{F}{E}\int_{-\infty}^t dt'\exp[i\bfk\cdot\bfomega (t'-t)]
	\Delta H_{\bfk}(\bfI,t').
\label{eq:rhos}
\end{equation}

	To determine the response density and potential, we first
introduce biorthonormal expansions (e.g. Kalnajs 1977; Weinberg 1989)
for the density,
\begin{equation}
\rho^s=\sum_{\ell m j} a_j^{\ell m}(t) d_j^{\ell m}(r) 
	Y_{\ell m}({\bf\Omega}),
\end{equation}
and potential,
\begin{equation}
\Phi^s=\sum_{\ell m j} a_j^{\ell m}(t) u_j^{\ell m}(r) 
	Y_{\ell m}({\bf\Omega}).
\end{equation}
The perturbing potential $\Phi^e$ has an analogous expansion in terms
of coefficients denoted $b_j^{\ell m}(t)$.  The condition of
biorthonormality relates the basis functions through Poisson's
equation
\begin{equation}
\nabla^2 u^{\ell m}_i=4\pi G d_i^{\ell m},
\end{equation}
and imposes orthogonality,
\begin{equation}
-{1\over 4\pi G}\int dr r^2 u^{* \ell m}_j d^{\ell m}_i=\delta_{ij}.
\end{equation}

	To substitute into equation (\ref{eq:rhos}), we first rewrite
these expansions in action-angle variables using expressions from
Tremaine \& Weinberg (1984).  The coefficient
\begin{equation}
\Delta H_{\bfk}=\sum_{\ell}\bigl[a_j^{\ell k_3}(t)+b_j^{\ell k_3}(t)\bigr]
	\biggl[W_{\bfk}^{\ell j}(\bfI)+
	{4\pi\over 3}p_j^{\ell m} X_{\bfk}(\bfI)\delta_{\ell 1}\biggl]
		V_{\ell k_2 k_3}(\beta),
\end{equation}
where $\cos\beta=I_3/I_2$ and $V_{\ell k_2 k_3}(\beta)$ is defined in
terms of rotation matrices.  The second term in brackets on the
right-hand side arises from the indirect potential.  The Fourier
coefficients have the definitions
\begin{equation}
W_{\bfk}^{\ell j}(\bfI)={1\over 2\pi}
	\int_{-\pi}^{\pi}dw_1\exp(-i k_1 w_1) u_j^{\ell k_3}
	\exp[ik_2(\psi-w_2)],
\end{equation}
and 
\begin{equation}
X_{\bfk}(\bfI)={1\over 2\pi}\int_{-\pi}^{\pi}dw_1\exp(-i k_1 w_1) r
	\exp[ik_2(\psi-w_2)].
\end{equation}
The angle $\psi-w_2$ has the definition
\begin{equation}
\psi-w_2=\int_{r_p(\bfI)}^{r(\bfI,w_1)}{dr(\omega_2-I_2/r^2)\over
	\sqrt{2[E-\Phi(r)]-I_2^2/r^2}}.
\end{equation}
Lastly,
\begin{equation}
p_j^{1 m}=\int dr r^2 d_j^{1 m} \pdrv{\Phi_0}{r}.
\end{equation}

	Now substituting these expansions into equation
(\ref{eq:rhos}), taking the inner product with respect to $u_i^{\ell m
*} Y_{\ell m}^*(\bfOmega)$, where
\begin{equation}
u_i^{\ell m *} Y_{\ell m}^*(\bfOmega)=
	\sum_{\bfk'} V^*_{\ell k_2 k_3}(\beta) W_{\bfk}^{\ell i *}({\bf I})
	\exp(-i{\bf k'}\cdot{\bf w}),
\end{equation}
and noting that $d{\bf r}d{\bf v}=d{\bf I}d{\bf w}$, we obtain a set
of coupled integral equations for the expansion coefficients:
\begin{eqnarray}
a_i^{\ell m}(t)&=& -{1\over 4\pi G}\int d\bfI d\bfw \drv{F}{E}
	\sum_{\ell m i \bfk\bfk'} (i\bfk\cdot\bfomega)
	\int d\tau \exp[i\bfk\cdot\bfomega(\tau-t)]
	[a_j^{\ell m}(\tau)+ b_j^{\ell m}(\tau)]
	\nonumber \\
       &&V^*_{\ell k_2' k_3'}(\beta) V_{\ell k_2 k_3}(\beta)
	W_{\bfk'}^{\ell i *}(\bfI) 
	\biggl[W_{\bfk}^{\ell j}(\bfI)+
	{4\pi\over 3}p_j^{\ell m} X_{\bfk}(\bfI)\delta_{\ell 1}\biggl]
	\exp[i(\bfk-\bfk')\cdot\bfw]
\label{eq:aj}
\end{eqnarray} 
Integrating over angles yields $(2\pi)^3\delta_{\bfk \bfk'}$.  To
integrate over $\beta$, we employ the orthogonality relation of
rotation matrices (Edmonds 1960):
\begin{equation}
\int d\cos\beta V^*_{\ell' k_2 k_3}(\beta)V_{\ell k_2 k_3}(\beta)=
	{2\over 2\ell +1}\vert Y_{\ell k_2}(\half\pi,0)\vert^2
	\delta_{\ell' \ell}\equiv C_{\ell k_2}\delta_{\ell'\ell},
\end{equation}
where
\begin{eqnarray}
C_{\ell k_2}&=&{2^{2k_2-1}\over \pi^2}
	{(\ell-k_2)!\over(\ell+k_2)!}\left[{\Gamma[\half(\ell+k_2+1)]
	\over\Gamma[\half(\ell-k_2)+1]}\right]^2, \ell+k_2 \quad{\rm even} 
	\nonumber \\
	&=&0 \qquad\qquad\qquad\qquad\qquad\qquad\qquad\qquad,
	\ell+k_2 \quad{\rm odd}.
\end{eqnarray}

	Changing variables from $I_1$ to $E$, where
$dI_1=dE/\omega_1$, we define the matrix kernel
\begin{equation}
K^{\ell m}_{ij}(\tau-t)= -{(2\pi)^3\over 4\pi G}\int dE{JdJ\over\omega_1}
	\drv{F}{E} \sum_{\bf k} C_{\ell k_2} 
	(i\bfk\cdot\bfomega)\exp[i\bfk\cdot\bfomega(\tau-t)]
	W_{\bf k}^{\ell i}\biggl[W_{\bfk}^{\ell j}+
	{4\pi\over 3}\delta_{\ell 1}p_j^{\ell m} X_{\bfk}\biggl] .
\end{equation}
This reduces equation (\ref{eq:aj}) to the form
\begin{equation}
a^{\ell m}_i(t)=\sum_j\int_{-\infty}^td\tau K^{\ell m}_{ij}(\tau-t)
[a^{\ell m}_j(\tau)+b^{\ell m}_j(\tau)].
\label{eq:aiinteq}
\end{equation}
The equations can be solved iteratively (Tricomi 1957).

	To derive the matrix equation in the frequency domain
(Weinberg 1989), we take the two-sided Laplace transform of equation
(\ref{eq:aiinteq}), so that
\begin{equation}
\tilde{\bf a}(s)=\tilde{\bf M}(s)[\tilde{\bf a}(s)+\tilde{\bf b}(s)],
\label{eq:matsol}
\end{equation}
where the matrix
\begin{equation}
\tilde M_{ij}^{\ell m}(s)=
	-{(2\pi)^3\over 4\pi G}\int
	dE{JdJ\over\omega_1}\drv{F}{E}\sum_{\bfk} C_{\ell k_2}
	{i\bfk\cdot\bfomega\over s+i\bfk\cdot\bfomega} W_{\bfk}^{\ell
	i *} \biggl[W_{\bfk}^{\ell j}+
	{4\pi\over 3}\delta_{\ell 1}p_j^{\ell m} X_{\bfk}\biggl].
\end{equation}

\subsection{Green's function and linear response operator}
\label{sec:gf}

	Define the Green's function for the density response in the
following way:
\begin{equation}
\rho^s(\bfr,t)=\int d\bfr' dt' G_{\rho}(\bfr,\bfr';t-t') \rho^e(\bfr',t').
\label{eq:Gdef}
\end{equation}
$G_{\rho}$ depends only on $t-t'$ because the system is
translationally invariant in time.  The notation $G_{\rho}$ indicates
that the Green's function determines the response density since we can
define an analogous relation for the response potential.  Introducing
a point source density
\begin{equation}
\rho^e(\bfr',t')=\delta(\bfr'-\bfr_p)\delta(t'-t_p),
\end{equation}
yields an equation for the Green's function
\begin{equation}
\rho^s_{\bfr_p,t_p}(\bfr,t)=G_{\rho}(\bfr,\bfr_p;t-t_p),
\end{equation}
where the notation $\rho_{\bfr_p,t_p}$ indicates that the response is
due to a point source at a particular location.  

	The linear response operator has a definition similar to that
of $G_{\rho}$ (Nelson \& Tremaine 1997):
\begin{equation}
\rho^s(\bfr,t)=\int d\bfr' dt' R(\bfr,\bfr';t-t') \Phi^e(\bfr',t').
\label{eq:Rdef}
\end{equation}
Comparing equations (\ref{eq:Gdef}) and (\ref{eq:Rdef}) shows that the
Green's function and linear response operator have the following
relation:
\begin{equation}
R(\bfr,\bfr';t-t')=G_{\rho}(\bfr,\bfr';t-t')(4\pi G)^{-1}\nabla^2_{\bfr'}
\end{equation}
The method defined above provides a numerical definition of the
Green's function or linear response operator for a system.

\section{Numerical implementation and tests}
\label{sec:implementation}
We use standard numerical techniques to implement the method.  There
are three basic components: the action-angle expansions, the basis
function expansions and the solution of the coupled integral
equations.  Angles and frequencies are computed from the orbit
equations on a $120\times 20$ grid in $E$ and $J$, respectively, with
$k_1\leq 4$.  Calculations with $k_1\leq 8$ produced the same results:
higher-order resonance influence damping at small spatial scales
rather than the large scales of interest here.  We use the Hernquist
basis for the biorthogonal expansion and truncate the series after 30
radial terms.  The coupled integral equations are solved iteratively
to a tolerance of $10^{-6}$ using Simpson's rule on a 1024-point grid
to evaluate the lag time quadrature in equation (\ref{eq:aiinteq}) for
$\tau\geq t-200$.

In practice, we first compute the angles, frequencies and Fourier
coefficients of the basis functions for a particular model and $\ell$
and dump the results to a file using an unformatted output stream.
Taking this data as input, we then compute the response kernel in
equation (\ref{eq:aiinteq}) and again save the results using an
unformatted output stream.  Computing the kernel takes the most time--
roughly 3 hours on a fully loaded SGI origin.  However, given the
kernel, we can calculate the response to any perturbation in short
order.

We performed three tests of the implementation.  The simplest was to
reproduce the analytic adiabatic fluid response calculation for both
the isothermal sphere (Murali \& Tremaine 1997) and a $W_0=5$ King
model.  This provides a test of the spatial response which is
independent of the temporal response.  It ensures that the basis
function expansion works properly and that a sufficient number of
terms have been retained to correctly reproduce the potential and
density response.  We then tested the accuracy of the spatial and
temporal solution by Laplace transforming the response coefficients
along a strip in the complex plane and plugging these values into the
dispersion relation, equation (\ref{eq:matsol})\footnote{Suggested by
John Dubinski}.  The transforms of the dominant coefficients typically
satisfy the dispersion relation to within at least 10\%.

	Finally, as a straightforward but important check, we compare
in Figure 1 the results of our linear calculation with an N-body
simulation performed with the self-consistent-field code discussed
most recently by Hernquist \& Ostriker (1992).  The code employs the
identical multipole-biorthonormal decomposition of the potential used
in the linear calculation developed above.  Thus we may directly
compare the time evolution of expansion coefficients computed using
the two methods.  We find good agreement for $\ell=2$.

\begin{figure}
\label{fig:disk.coef}
\plotone{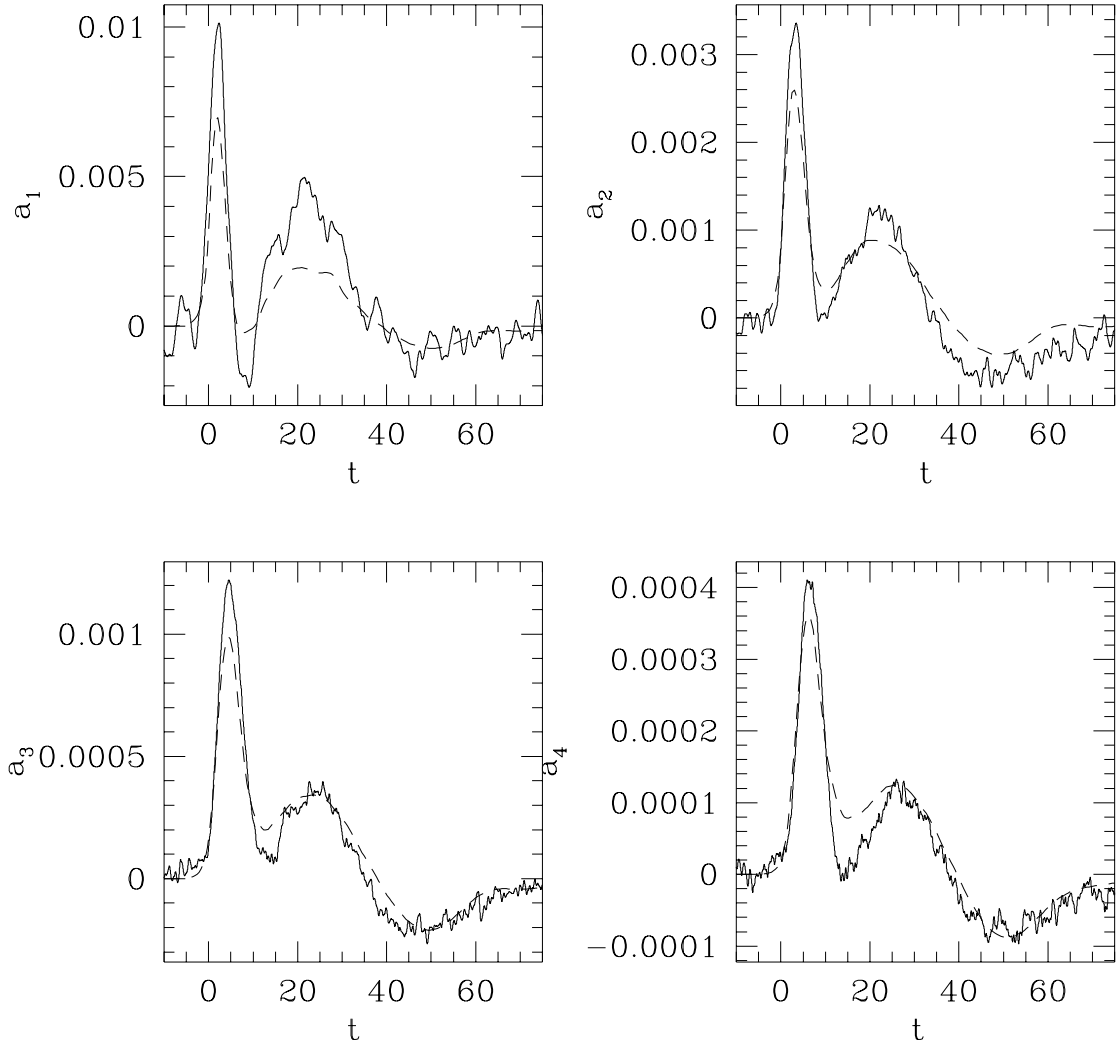}
\caption{The self-gravitating response of a $W_0=5$ King model to an
$\ell=2$, $m=0$ perturbation which is turned on and off exponentially.
The four lowest order expansion coefficients are plotted in time for
the linear calculation (dashed) and N-body simulation (solid) with
$10^6$ particles.}
\end{figure}	

Given the results below, we are particularly interested in comparing
the two calculations for $\ell=1$.  However, we were unable to
reproduce the calculated response using an SCF simulation with $10^6$
particles.  As Weinberg (1994) notes, expansion codes do not conserve
momentum, so they will be hard-pressed to reproduce an off-center
response since the center wanders.  In particular, to defeat
wandering, the standard centering algorithm calls for resetting the
coordinate center to the center-of-mass of the system at each
potential recomputation.  However, to recognize and reproduce a dipole
response we should ideally remain in a reference frame which is
attached to the center of the unperturbed potential (which is no
longer at the center-of-mass because of the dipole response).
Nevertheless, experiments without the centering algorithm also failed
to reproduce the effect.

\section{Response to point-like perturbations}
\label{sec:point_perturber}
In the spirit of calculating the Green's function, we subject the
model systems to point-like perturbations:
\begin{equation}
\Phi^e(\bfr)=\delta(t)\Phi_p(\vert\bfr-\bfr_p\vert)
\end{equation}
where $\Phi_p(\bfr_p)$ denotes the potential of a point-mass of mass
$m_p$ located at $\bfr_p$ (for definiteness, the angular coordinates
$\theta_p=\phi_p=0$).  Point-like perturbations also approximate
Poisson fluctuations in equilibrium systems.

We impose perturbations of harmonic order $\ell=0,1,2$ and
corresponding $m$ values on a range of King models at radii enclosing
different mass fractions of the unperturbed system.  Table 1 outlines
the computational grid.  The notation $r_x$ denotes the radius
enclosing the fraction $x$ of the total mass of the system.  The
quantity $r_t$ denotes the tidal radius or radius enclosing 100\% of
the mass.  

Units are defined such that the total mass $M=1$, $G=1$ and the total
energy $E=-1/8$.  This implies that, at $r_{1/2}$, the period of a
circular orbit, $P_{1/2}\sim 18$.  Reaonable physical values can be
obtained by rescaling the models to halos of mass $M=10^{12}
M_{\odot}$ and limiting radius $r_{max}=200 \kpc$.

\begin{table*}
\label{tab:km}
\caption{Characteristics of King models and computational grid}
\begin{tabular}{lrrrrr}
\\
$W_0$&$c$&$r_{1/4}$&$r_{1/2}$&$r_{3/4}$&$r_t$\\
3&0.7&1.1&1.7&2.4&6.2\\
5&1.0&1.0&1.6&2.6&8.7\\
7&1.5&0.8&1.6&3.2&14\\
\end{tabular}
\end{table*}

We set the perturbation amplitude by adjusting $m_p$.  Our choices are
arbitrary and serve more to illustrate the physics of the response.
For distant perturbations $M(r_p)>0.75$, we typically choose $m_p=0.1$
while for inner perturbations $M(r_p)<0.75$, we typically choose
$m_p=0.01$.  Since the calculation is linear, we may rescale the
results for other choices of $m_p$.

\subsection{Transmission and amplification of disturbances}
We find that significant transmission and amplification of
disturbances can occur in the model calculations.  Since the response
to each $\ell$ is independent, we consider each separately.

\subsubsection{Monopole response}
The monopole component of the perturbation acts like a spherical shell
of mass $m_p$ which exerts a force only on material at $r>r_p$, the
shell radius.  This generates a density enhancement at $r>r_p$ which
propagates inward to smaller radii.  Figure 2 shows this effect in a
$W_0=3$ King model.  We find that the amplitude of the propagating
response can be significant well within the radius of the
perturbation.  Since the monopole component of the perturber's direct
gravitational field exerts no force on material with $r<r_p$, the
amplification is infinite.

\begin{figure}
\label{fig:w3.l0.4.trans}
\plotone{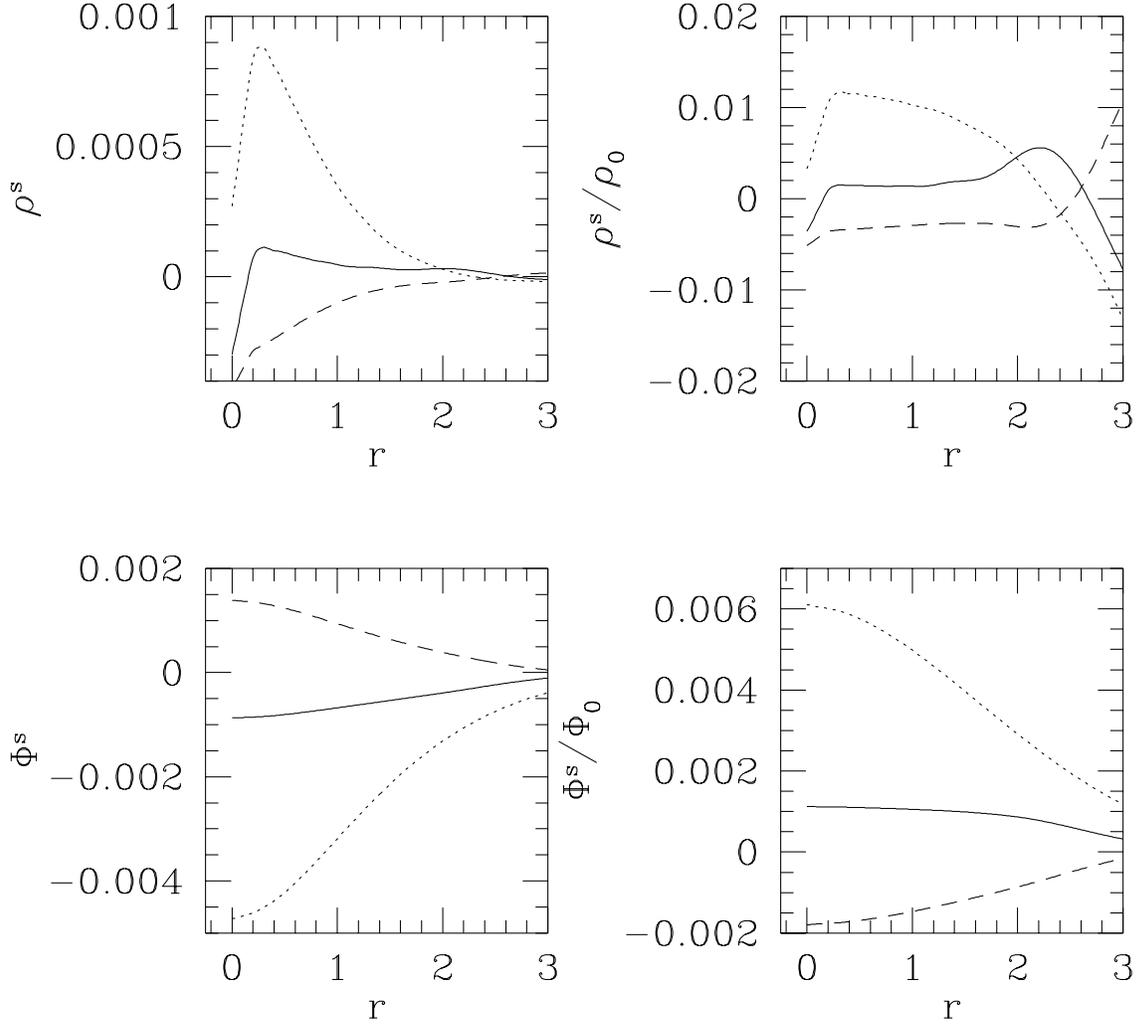}
\caption{Transmission of spherical disturbance in $W_0=3$ King model
initially excited at $r_p=2.4$ (75\% mass radius) for $m_p=0.1$.
Panels show density, potential and respective contrasts as a function
of radius as indicated.  In each panel, solid line shows inwardly
growing response at $t=1.1$ following perturbation.  Dotted line shows
peak central response at $t=5.8$.  Dashed line shows subsequent
minimum at $t=12.8$ as disturbance propagates outwards.  The
perturbation subsequently damps away.  The peak central density
response is roughly 1\%.}
\end{figure}

Deeper within the $W_0=3$ system, smaller perturbations give rise to
similar effects in the core.  For perturbations at the 25\% mass
fraction, we find a roughly a 1\% central response for $m_p=0.01$.
The amplitude of the peak central density response decreases with
concentration due to the increasing depth of the central potential
well.

\subsubsection{Dipole response}
The dipole component of the perturbation acts somewhat differently
from the monopole in that it exerts a force at all radii.  Some
component of this force contributes to the mean motion of the system;
recall that the indirect potential removes this component in the
present calculation.  Nevertheless, the residual (differential)
acceleration excites a strong dipolar response in all the models
studied here.

For each of the King models, Table \ref{tab:peak75} shows the basic
response characteristics for perturbations at the 75\% mass radius.
The response travels deep into the system, producing peak density
contrast at roughly $0.1 r_p$ (radius denoted $r_{peak}$ enclosing
$M(r_{peak})$ of the total mass).  In each case, the fractional
density response is several \%.  The response improves with
concentration.  Also note that there may be a substantial delay
between the application of the perturbation and the peak response in
the system's core at $t_{peak}$.

\begin{table*}
\caption{Peak central density contrast for $M(r_p)=0.75$ and
$m_p=0.1$}
\label{tab:peak75}
\begin{tabular}{lrrrrr}
\\
$W_0$&$r_p$&$r_{peak}$&$\vert\rho^s/\rho_0\vert$&$M(r_{peak})$&$t_{peak}$\\
3&2.4&0.2&0.03&0.01&5.8\\
5&2.6&0.3&0.04&0.02&5.8\\
7&3.2&0.3&0.05&0.03&8.1\\
\end{tabular}
\end{table*}

Figure 3 provides a close look at a cross-section of the response in
the $W_0=7$ system.  The density at small radii responds strongly to
the perturbation.  To better understand this effect, we show the cross
section of the mass perturbation $r^2\rho^s$ along $\bfr_p$ in Figure
4.  The perturbation draws mass from all surrounding directions.
However, mass drawn from inner regions produces a large change in the
density because of the smaller volume.

\begin{figure}
\label{fig:w7.l1.4.trans}
\plotone{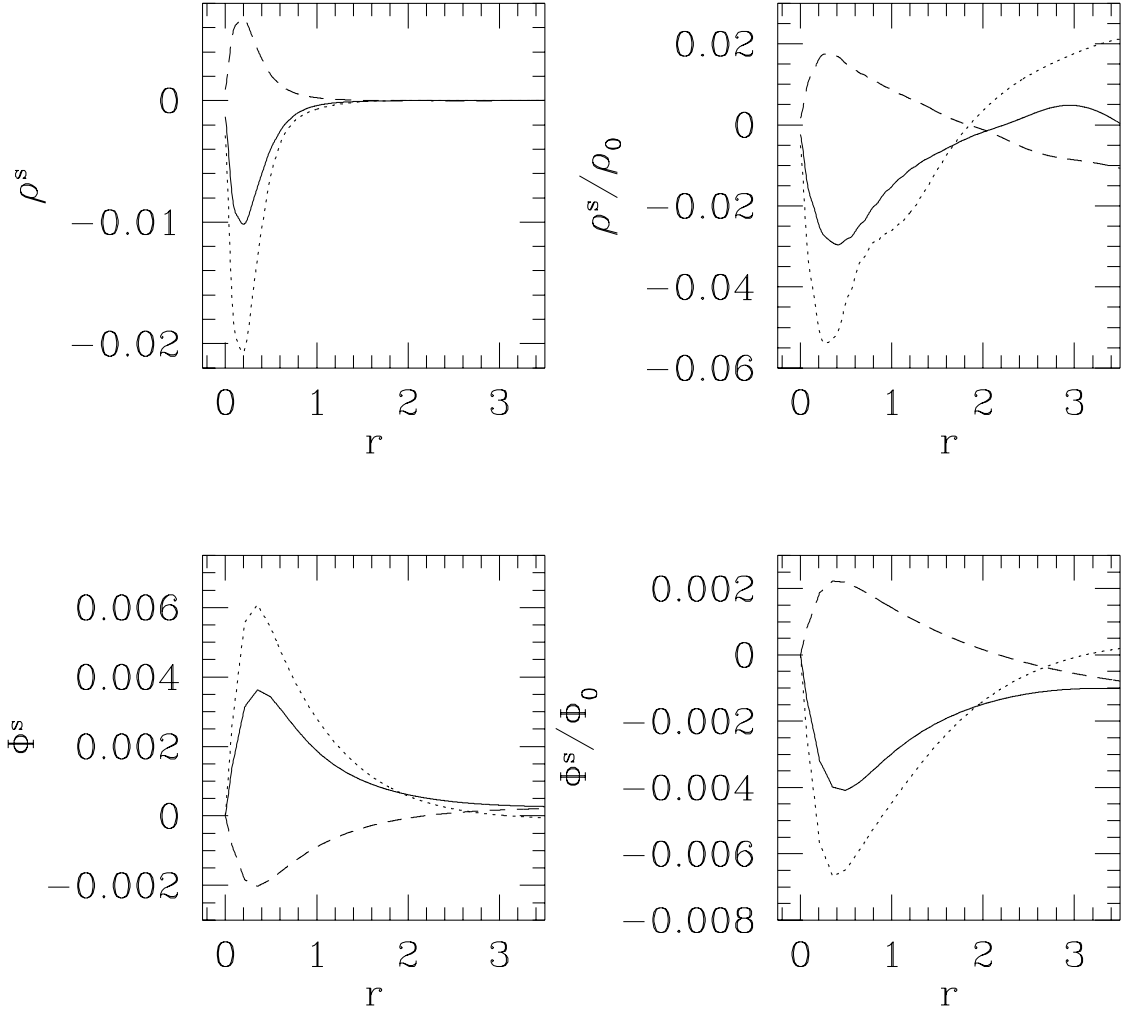}
\caption{Transmission of dipole disturbance in $W_0=7$ King model
initially excited at $r_p=3.2$ (75\% mass radius) for $m_p=0.1$.
Panels are as in previous figure with radial profile along $\bfr_p$.
Solid line shows growing response at $t=3.4$ following perturbation.
Dotted line shows peak central response at $t=8.1$.  Dashed line shows
large subsequent peak after next half-cycle of mode oscillation at
$t=38.6$, indicating the presence of weakly damped modes with very
slow pattern speeds.  The peak inner density response is roughly 5\%,
dropping to $\sim 2\%$ after the next half-cycle.}
\end{figure}

\begin{figure}
\label{fig:w7.l1.4.dm}
\plotone{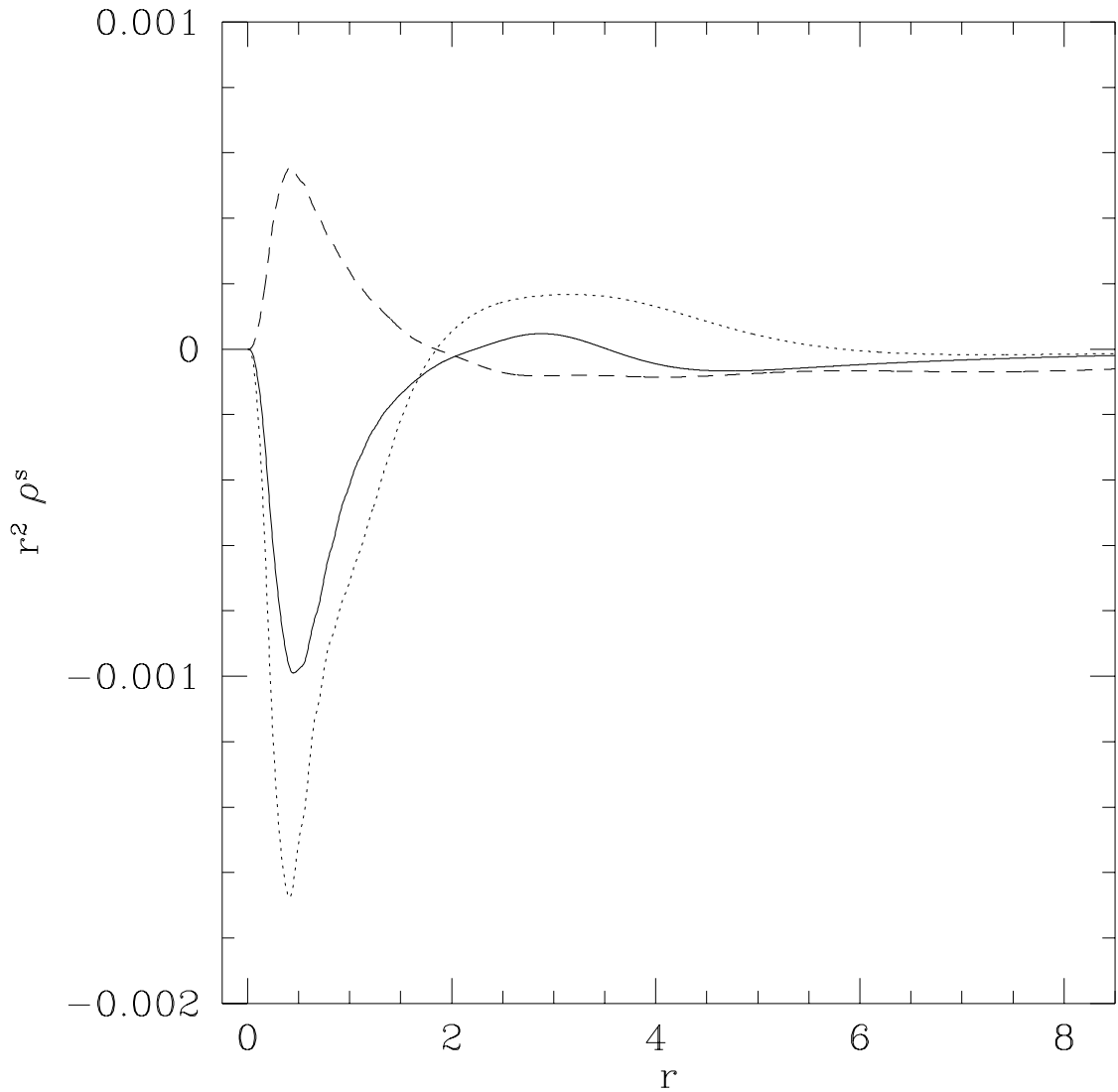}
\caption{The change in mass along the $\bfr_p$ axis corresponding to
the perturbation depicted in the previous figure.  At $t=3.4$, mass is
attracted towards the perturber's original position ($r_p=3.2$;
solid).  This draws material from surrounding regions, but affects the
density small radius because of the smaller volume.  At $t=8.1$, the
response peaks (dotted).  Finally, after the next half-cycle
($t=38.6$), the response peaks again (dashed).}
\end{figure}

To gauge the importance of self-gravity, it is useful to compute the
amplification factor or Love number of the perturbation:
$\chi\equiv(\Phi^s+\Phi^{e\prime})/\Phi^{e\prime}$, where
$\Phi^{e\prime}$ includes the indirect term\footnote{There is
apparently some ambiguity in defining $\chi$ for $\ell=1$: we have
chosen $\Phi^s$ as the total potential perturbation and
$\Phi^{e\prime}$ as the external potential which produces $\Phi^s$.}.
Figure 5 shows $\chi$ for the perturbation depicted in the previous
figure.  The $W_0=7$ system achieves dipole amplification factors of
10 or so which may persist for long durations in weakly damped modes.
Such strong amplification will lead to non-linear effects that are not
described by our present treatment.  The amplification factors
decrease with concentration: for $W_0=5$, $\vert\chi\vert\sim 4$; for
$W_0=3$, $\vert\chi\vert\sim 2$.

\begin{figure}
\label{fig:w7.l1.4.amp}
\plotone{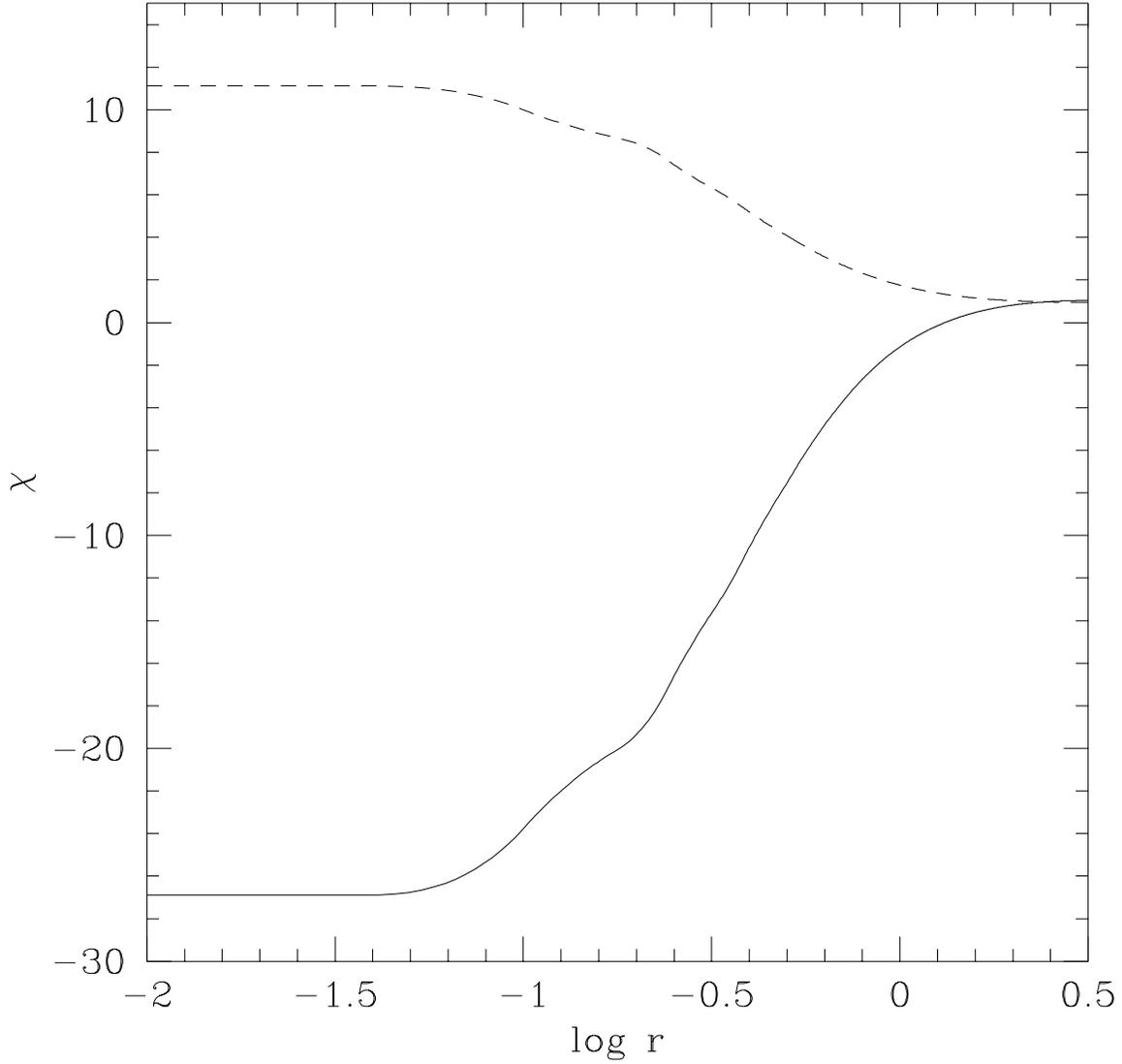}
\caption{Potential amplification factor $\chi$ along $\bfr_p$ for
disturbance shown in figure \ref{fig:w7.l1.4.trans} ($\log r_p=0.5$).
Solid line shows amplification at peak response ($t=8.1$).  Dashed
line shows amplification at peak after next half-cycle ($t=38.6$).  On
the near side of the disturbance, de-amplification of the potential
perturbation weakens the interior halo potential while, on the far
side, de-amplification strengthens.  The amplification magnitude is
roughly 10 and can persist for long times after the disturbance
disappears, due to the excitation of weakly damped modes.}
\end{figure}

Similar effects result from disturbances at smaller radius.  These
lead to stronger effects in the core.  Table \ref{tab:peak50} shows
the peak central response arising from a point perturbation at the
radius enclosing 50\% of the mass.  Here disturbances are transmitted
to small radius.  Disturbances at even smaller radius can excite the
dominant response directly.

\begin{table*}
\caption{Peak central density contrast for $M(r_p)=0.50$ and
	$m_p=0.01$}
\label{tab:peak50}
\begin{tabular}{lrrrrr}
\\
$W_0$&$r_p$&$r_{peak}$&$\vert\rho^s/\rho_0\vert$&$M(r_{peak})$&$t_{peak}$\\
3&1.6&0.7&0.004&0.09&17.5\\
5&1.6&0.6&0.006&0.09&15.2\\
7&1.6&0.3&0.009&0.03&20.0\\
\end{tabular}
\end{table*}

\subsubsection{Quadrupole response}
In these experiments, quadrupole disturbances do not appear to
propagate from large radius to small radius in the same manner as
monopole and quadrupole disturbances.  Figure 6 shows an example of
the response to an $\ell=2$, $m=2$ perturbation in a $W_0=5$ King
model.  The main feature is the local density response which then
decays away relatively rapidly.  This tends to produce moderate
amplifications of $\sim 30\%$ at radii roughly $0.3r_p$.  Overall, the
quadrupole response appears to be most important in the case of pure
resonant forcing (Weinberg 1998).

\begin{figure}
\label{fig:w5.l2.2.trans}
\plotone{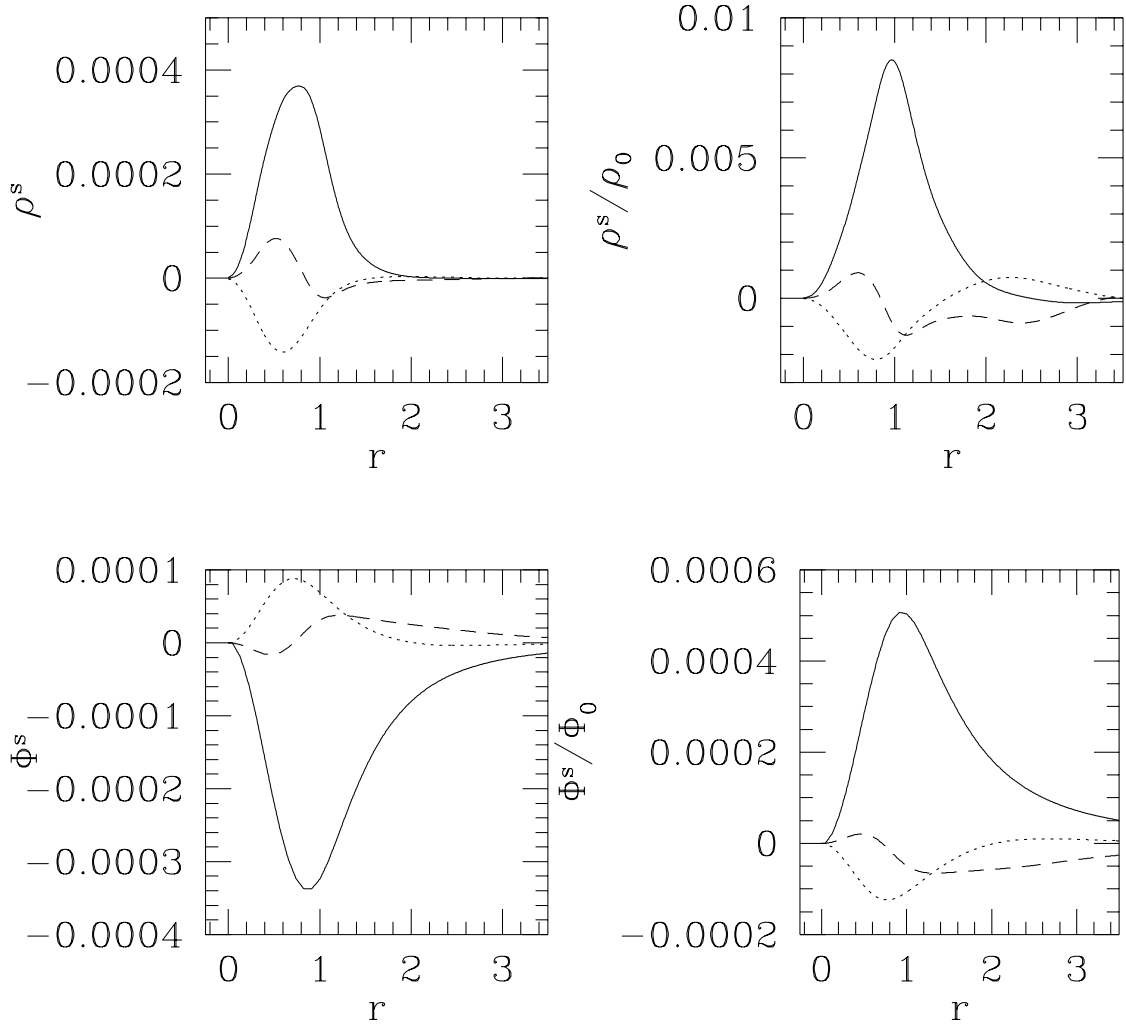}
\caption{Transmission of quadrupole response in $W_0=5$ King model
with $r_p=1.0$ (25\% mass radius) for $m_p=0.01$.  Panels are as in
above figures with radial profile along $\bfr_p$.  Solid line shows
peak response at $t=1.1$, immediately after perturbation.  Dotted line
and dashed lines show subsequent decay at $t=3.4$ and $t=5.7$,
respectively.}
\end{figure}

\section{A realistic example of dipole response}
\label{sec:realistic}
The above analysis details the physical behavior of the response.  To
provide a more realistic example focusing in the $\ell=1$ response, we
take a point-mass perturber with $m_p=0.01$ on a hyperbolic trajectory
through a $W_0=7$ halo.  The perturber initially has velocity $v=0.5$
at infinity: this is roughly the circular speed at the half-mass
radius.  Closest approach is at $r=2.6$, the radius enclosing 68\% of
the total mass.  The maximum velocity is roughly $v=1.0$ so that the
effect of dynamical friction will be small.  Figure 7 shows the strong
$\ell=1$ response induced by the relatively small perturber.

\begin{figure}
\label{fig:flyby}
\plotone{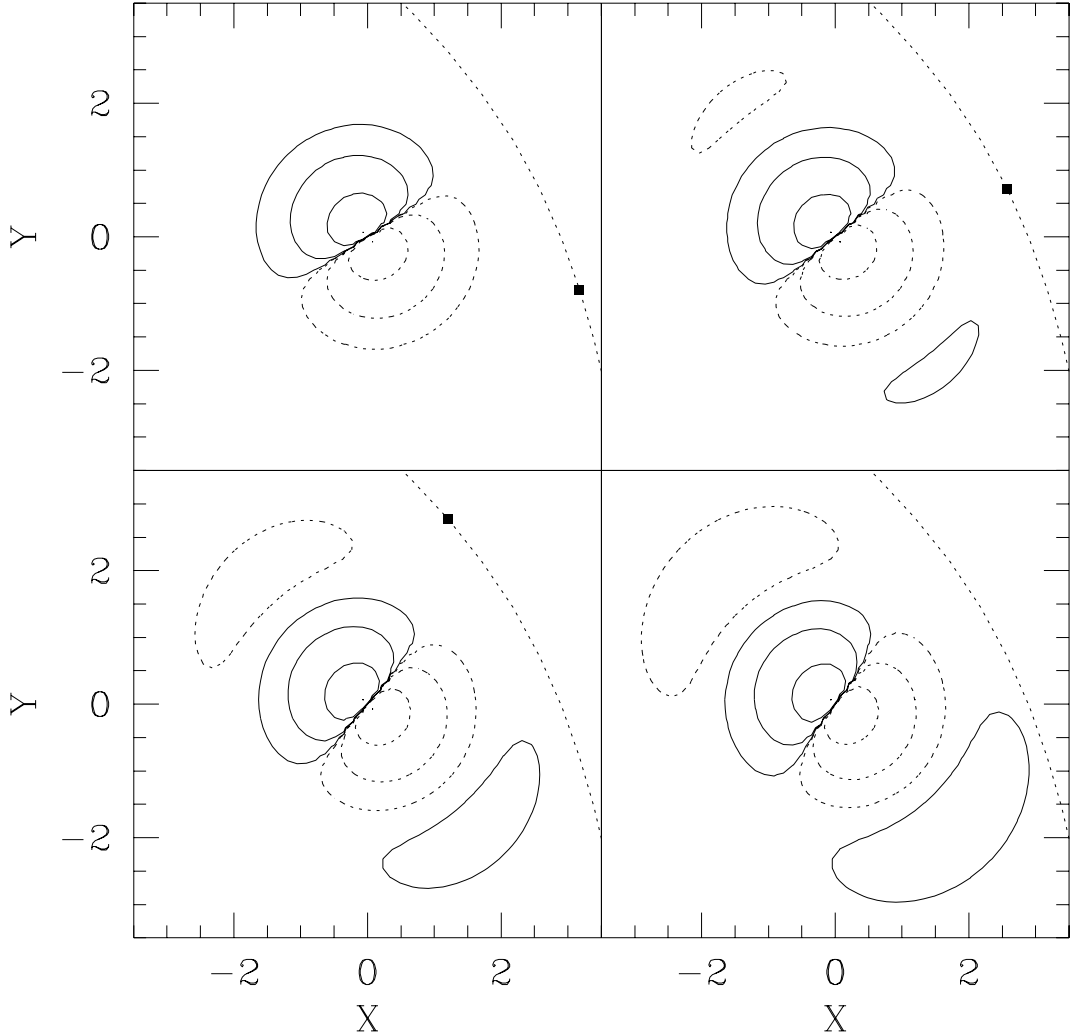}
\caption{The dipole response induced by a point-mass perturber with
$m_p=0.01$ on a hyperbolic trajectory in the X-Y plane indicated by
dotted curve in each panel.  The solid square shows the perturber's
position when in the field of view.  Solid (dotted) contours show
overdensity (underdensity) in unit logarithmic intervals,
$\log\rho^s=-1,-2,-3$, about the peak at each time.  Time interval
$\Delta t=2.44$ separates each panel.  The peak density response is
$\rho^s_{max}=0.022, 0.032, 0.039, 0.042$ at $r_p=0.2$ from left to
right, top to bottom.  The absolute maximum occurs in the final panel
after the perturber has flown by and is roughly a 10\% perturbation to
the background density.  The mode continues to librate through the
center long after the perturber has passed.}
\end{figure}

The encounter excites a very weakly damped mode.  Figure 8 shows the
time-dependence of the induced potential energy perturbation which
provides a good signature of mode oscillation.  For comparison,
Weinberg (1994) found an e-folding $2\pi/\Im(\omega)=250$ and an
oscillation period $2\pi/\Re(\omega)=98$ in these units for $W_0=7$.
The response depicted here shows an oscillation period $P\approx 100$
in good agreement.  The damping rate is harder to judge because the
apparent amplitude consists of a complex superposition of interfering
modes; nevertheless, the damping rate is low and appears to have
roughly the same magnitude.  Also recall that the precise character of
the dipole modes will differ because of the difference in Hamiltonian.

\begin{figure}
\label{fig:w7.bh.l1.2.w}
\plotone{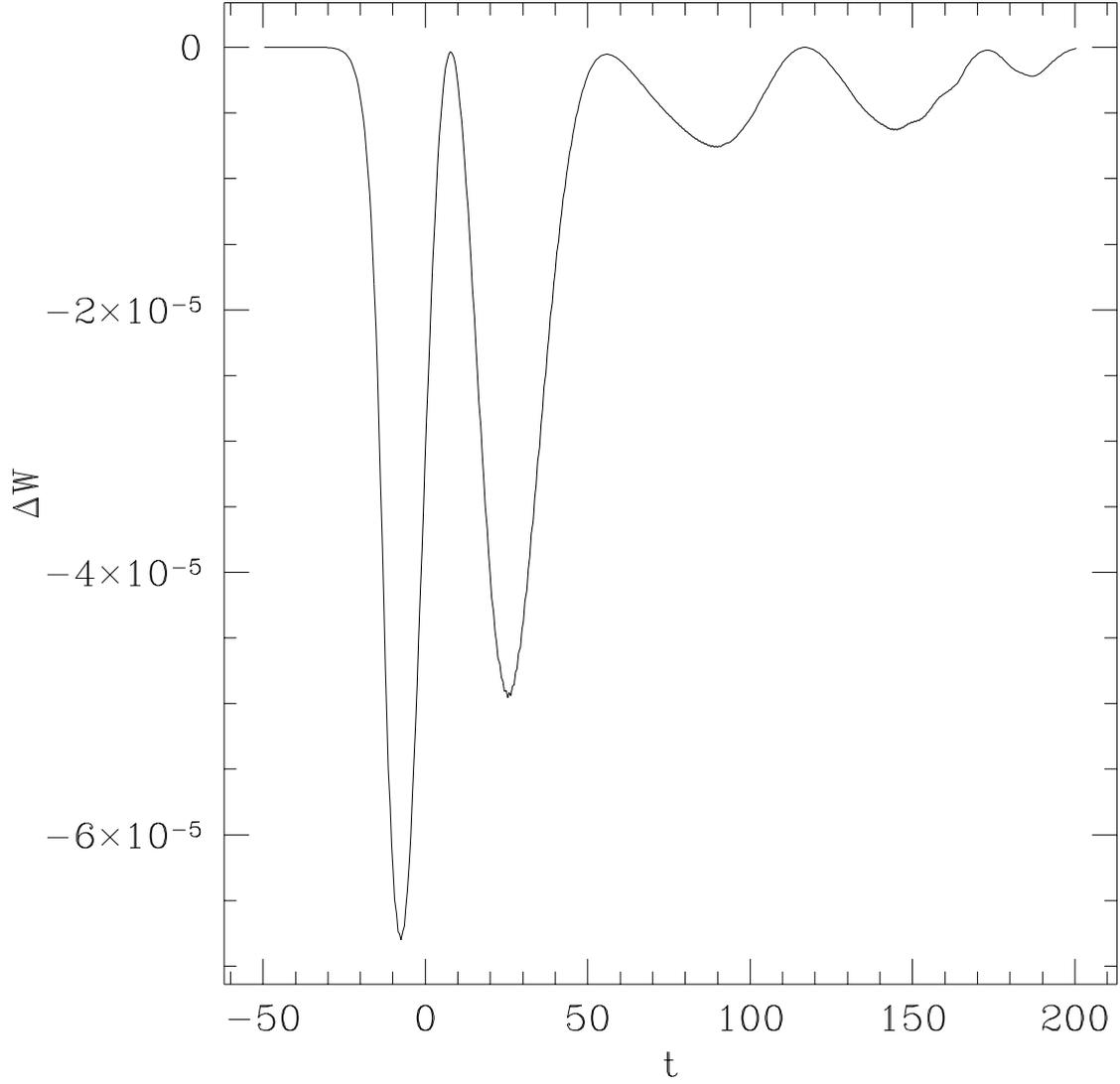}
\caption{Potential energy perturbation as a function of time for
$W_0=7$ system in Figure \ref{fig:flyby}.  The potential energy
perturbation is only $\sim3\times10^{-4}$ the potential energy of the
entire system.  The total initial disturbance decays rapidly but a
fair amount of the input power persists in a weakly damped mode.}
\end{figure}

\section{Discussion and Conclusions}
\label{sec:discussion}
Overall, our results indicate that halos can respond quite actively to
perturbations.  The response appears strong over a wide range of
profiles and can efficiently transmit disturbances from the outer halo
into the inner halo.  In particular, the dipole response dominates and
is strongest for the highest concentration system.  The weak damping
at high concentration may reflect kinematic decoupling of core and
halo.  The dipole response is still significant at low concentration
but does not appear to persist in weakly damped modes.  At low
concentration, the monopole contribution can be fairly strong and
long-range.  The quadrupole response is roughly independent of
concentration and seems to be more strongly excited through resonant
forcing (Weinberg 1998).

Our results complement recent work by Weinberg (1994).  The results
appear to be consistent: mode shapes are similar; both calculations
give roughly the same central displacement due to the dipole response;
and the weakly damped dipole modes in the $W_0=7$ system appear to
have nearly the same period and similar damping rates in both
calculations.  Furthermore, the present results suggest that it may be
difficult to excite the weakly damped dipole modes in $W_0=3$ and
$W_0=5$ King models discussed by Weinberg (1994).

More generally, the results suggest that noisy halos play a role in
exciting disk structure and can drive evolution in the inner parts of
galaxies.  As an example, recent surveys of disk galaxies show a high
frequency of lopsided galactic disks ($\sim 30\%$; Zaritsky \& Rix
1997).  The strong $\ell=1$ response found in our calculations
suggests that the halo has influence in these systems.  In particular,
a weakly damped $\ell=1$ mode with a long period can induce an
adiabatic $m=1$ distortion in a disk.  Scaling the $W_0=7$ results to
a 200 kpc halo, we find that the location of peak response is roughly
4 kpc, which is consistent with typically observed distortion
locations.  Persistent modes are attractive also because, in many
cases, there is no obvious companion to the galaxy in question.  Thus
a previous encounter can excite a long-lived mode that continues to
slowly distort the disk long after the perturber has gone.  Ultimately
the disk response depends on the relative mass in the mode.  This
aspect will require further investigation.

Another significant possibility is that active halos strongly
influence the formation of galactic disks.  The standard scenario is
that disks settle adiabatically into static dark halos (e.g. Dubinski
1994).  However, the proto-galactic environment is likely to be
extremely noisy, particularly in outer regions, so that the halo may
perturb the disk by transmitting numerous disturbances into the inner
galaxy.  Moreover, the inner halo may continue to oscillate as it
settles after initial collapse.  Halo oscillations can easily perturb
the disk through the time-dependent gravitational potential.
Conversely, the structural integrity of observed disks set limits on
the degree of disequilibrium in the proto-galactic halo.

The above calculations have proved extremely valuable because of their
high precision: they provide further evidence that the very subtle
influence of the halo plays a major role in determining the observed
properties of disk galaxies.  Nevertheless, we emphasize the need to
improve and perhaps develop alternatives to the methods used herein.
The main drawback of the above method is the reliance on numerous
expansions and the extreme care required in implementation.  Moreover,
we are limited to systems with spherical geometry whereas galaxies
most likely come in a variety of shapes.  However, commonly used
N-body methods still seem far from achieving the sensitivity required
to reproduce such subtle effects.  To reproduce off-center modes might
require direct N-body codes with at least $10^6$ particles-- still 1
to 2 orders of magnitude beyond current capabilities.

\begin{acknowledgments}
This research was supported by NSERC and the Fund for Astrophysical
Research.  I thank John Dubinski for discussion and for assistance in
setting up the parallel field code.  I also thank Phil Arras, Scott
Tremaine, Ira Wasserman, Martin Weinberg, Dick Bond and Peter
Goldreich for helpful discussion.
\end{acknowledgments}

\end{document}